\documentclass[AMA,STIX1COL]{WileyNJD-v2}
\usepackage{moreverb}
\usepackage{amsmath}

\newcommand\BibTeX{{\rmfamily B\kern-.05em \textsc{i\kern-.025em b}\kern-.08em
T\kern-.1667em\lower.7ex\hbox{E}\kern-.125emX}}

\received{<day> <Month>, <year>}
\revised{<day> <Month>, <year>}
\accepted{<day> <Month>, <year>}

\begin{document}

\title{Understanding Difference-in-differences methods to evaluate policy effects with staggered adoption: an application to Medicaid and HIV}

\author[1]{Julia C. Thome, B.A.*}

\author[2]{Peter F. Rebeiro, Ph.D.}

\author[2]{Andrew J. Spieker, Ph.D.}

\author[2]{Bryan E. Shepherd, Ph.D.}

\authormark{Thome \textsc{et al}}

\address[1]{ \orgname{Vanderbilt University}, \orgaddress{\city{Nashville}, \state{Tennessee}}}

\address[2]{\orgname{Vanderbilt University Medical Center}, \orgaddress{\city{Nashville}, \state{Tennessee}}}

\corres{*Julia C. Thome, \email{julia.c.thome@vanderbilt.edu}}

\presentaddress{2525 West End Ave, Nashville, TN 37203}

\abstract[Abstract]{While a randomized control trial is considered the gold standard for estimating causal treatment effects, there are many research settings in which randomization is infeasible or unethical. In such cases, researchers rely on analytical methods for observational data to explore causal relationships. Difference-in-differences (DID) is one such method that, most commonly, estimates a difference in some mean outcome in a group before and after the implementation of an intervention or policy and compares this with a control group followed over the same time (i.e., a group that did not implement the intervention or policy). Although DID modeling approaches have been gaining popularity in public health research, the majority of these approaches and their extensions are developed and shared within the economics literature. While extensions of DID modeling approaches may be straightforward to apply to observational data in any field, the complexities and assumptions involved in newer approaches are often misunderstood. In this paper, we focus on recent extensions of the DID method and their relationships to linear models in the setting of staggered treatment adoption over multiple years. We detail the identification and estimation of the average treatment effect among the treated using potential outcomes notation, highlighting the assumptions necessary to produce valid estimates. These concepts are described within the context of Medicaid expansion and retention in care among people living with HIV (PWH) in the United States. While each DID approach is potentially valid, understanding their different assumptions and choosing an appropriate method can have important implications for policy-makers, funders, and public health as a whole.}


\maketitle

\section{Introduction}\label{sec1}

\noindent A randomized control trial (RCT) is the gold standard for estimating causal treatment effects.\cite{Sibbald201} However, there are many research settings in which randomization is infeasible or unethical. Quasi-experimental research designs offer an alternative to RCTs in such settings and many analytical methods use data from these research designs to estimate causal relationships.\cite{shadish_experimental_2002}\cite{Campbell} The Difference-in-differences (DID) method is one such method that, most commonly, estimates a difference in some mean outcome of a group before and after an intervention or policy and compares this with a control group followed over the same time that did not experience the intervention or policy.\cite{Wing} The first recorded use of DID was by John Snow in 1855 to determine the water-borne cause of Cholera in London.\cite{Snow} Snow's analysis included two exposure groups (exposure to contaminated wells versus no exposure) and two time periods (pre-exposure versus post-exposure), which remains the most popular and well-understood DID setup used by researchers today.\\

\noindent As we continue to tailor DID approaches to accommodate complex scenarios encountered in the real world, we must consider how the underlying assumptions associated with those methods change. DID methods, as well as their extensions and applications, have been popular within the economics and econometrics literature and are often included in economic textbooks. \cite{gertler_wilde_premand_rawlings_vermeersch_2016}\cite{Angrist}\cite{ wooldridge_2021}\cite{white_raitzer_2017} Many overviews of these methods and extensions have been produced over the past two decades in the economics literature. \cite{fredriksson_oliveira_2019}\cite{Lechner2011}\cite{Angrist_Krueger}\cite{Bertrand}\cite{Blundell}\cite{Imbens_Wooldridge}\cite{Athey_Imbens}\cite{Abadie_Cattaneo} \cite{ROTH2023} \cite{dechaisemartin2022twoway} Recently, DID methods have been increasingly used for public health research;\cite{ryan_burgess_dimick_2014} their proper uses have been summarized at a high level in the literature.\cite{Wing}\cite{Dimick2014MethodsFE} \cite{ryan_burgess_dimick_2014} On the other hand, recent publications in the economics literature have highlighted potential issues with these methods in the context of more than two periods and two treatment groups. \cite{GoodmanBacon}\cite{deChaisemartin2020} DID modeling approaches have emerged in the economics literature that address these potential issues,\cite{Callaway} however, to our knowledge, these newer approaches are not mentioned in DID summaries found in the public health literature.\\ 

\noindent In this paper, we focus on extensions of the DID method and their relationship to linear models in the setting of staggered treatment adoption within the public health context. Throughout this manuscript, we utilize an application to Medicaid expansion and retention in care among people living with HIV (PWH) in the United States as an anchor for describing and presenting the methodology. We focus on the identification and estimation of the average effect of expansion on retention in care among PWH in expansion states using potential outcomes notation, highlighting the assumptions necessary to produce valid estimates using multiple DID extensions. While each approach is valid under its specific set of assumptions, understanding their different assumptions and choosing an appropriate method for addressing any specific policy-related question accordingly can have important implications for policy-makers, funders, and public health as a whole. We hope to shed light on recent extensions to DID methodologies and related modeling approaches within the context of public health research in order to provide guidance on the proper use and implementation in such contexts.\\

\section{Data and Notation}\label{data and notation}
\subsection{The Data}\label{data}

\noindent This study uses data from the North American AIDS Cohort Collaboration on Research and Design (NA-ACCORD), the largest and most geographically diverse HIV clinical cohort consortium in North America.\cite{NA-ACCORD} It comprises over 200 sites across the United States and Canada and has data on over 190,000 PWH. NA-ACCORD data between 2012 and 2017 are used in this paper.\\

\noindent The exposure of interest is Medicaid expansion between 2012-2017. Medicaid provides federal financial assistance to states within the United States operating approved medical-assistance plans and as of January 2017 was the nation's largest source of health coverage. \cite{Mazurenko} Historically, Medicaid has covered low-income children, parents, pregnant women, disabled people, and non-disabled adults without dependent children. In 2010, the Affordable Care Act expanded Medicaid coverage, extending eligibility to non-elderly adults with an income of up to 138\% of the federal poverty level. Medicaid was intended to be expanded in all states, however, a 2012 Supreme Court decision made expansion optional for states starting on January 1, 2014.\cite{Miller} \cite{Sommers}\cite{Mazurenko} By the end of 2020, 36 states (including DC) adopted Medicaid expansion: 26 in 2014, three in 2015, two in 2016, two in 2019, and three in 2020.\cite{kff_2021} The outcome of interest is retention in clinical care among PWH. Retention in care is defined as two or more completed visits more than 90 days apart during the given year. \cite{ford_spicer_2012}\cite{Rebeiro2014} Retention in care was assessed in each calendar year after the year of entry into the study. Covariates in the data include age, race, sex, and region. Age (represented by participant age at enrollment), race (non-Black or Black), and sex (Male or Female) are included in the data at the individual level. Participants' states of residence were grouped into US Census-designated regions at the state level: West (WE), Northeast (NE), South (SO), and Midwest (MW). \cite{CensusRegions}

\subsection{Notation}\label{notation}

\noindent We consider years $T = t$ where $t = 2012, \ldots, 2017$ and expansion groups $G = g$ where $g = 2014, 2015, 2016$, such that states in group $G = 2014$ expanded in 2014, states in group $G = 2015$ expanded in 2015, and so on. There were no states that expanded Medicaid in 2017.\cite{kff_2021} No states that expanded Medicaid subsequently reversed the expansion. States that did not expand by 2014, for example, are represented by $G > 2014$ and states in $G > 2017$ never expanded between 2012-2017.\\

\noindent We define $Y_{t}$ as the observed outcome of interest at time $T = t$: 1 if an individual was clinically retained during year $t$ and 0 otherwise. Let the expansion status of a state an individual lived in at time $t$ be represented by $Z_t = z$ where $z \in \{0,1\}$, such that $Z_t = 0$ indicates a state had not expanded Medicaid during year $t$ and $Z_t = 1$ indicates Medicaid was expanded during year $t$. Since the earliest expansion could occur was 2014, $Z_{2012} \equiv Z_{2013} \equiv 0$, but $Z_t$ could be equal to 0 or 1 for $t \geq 2014$. We can define the vector of expansion statuses associated with an individual up to time $t$ as $\overline{Z}_t = \left( Z_{2012}, \ldots, Z_t \right)$ . \\

\noindent We use the above to define the potential outcome notation that will be used in this paper. First, we define the potential outcome during year $t$ with hypothetical expansion status $z$ as $Y_t^{Z_t = z}$. For example, consider the potential outcome for an individual in 2016 under the hypothetical scenario in which Medicaid is expanded during 2016, $Y_{2016}^{Z_{2016} = 1}$.  In this simple case, we do not know if Medicaid first expanded in 2014, 2015, or 2016; we only know that it is currently expanded in 2016. To distinguish between expansion groups (i.e. the first year of expansion), we have to consider the potential outcome given a vector of hypothetical expansion statuses. For example, $Y_{2016}^{\overline{Z}_{2015} = \mathbf{0}, Z_{2016} = 1}$ represents the potential outcome in 2016 under the hypothetical scenario in which Medicaid first expanded in 2016 ($G = 2016$). In contrast, $Y_{2016}^{\overline{Z}_{2014} = 0, Z_{2015} = 1, Z_{2016} = 1}$ is the potential outcome in 2016 under the hypothetical scenario in which Medicaid first expanded in 2015 ($G = 2015$). Because a state remains expanded once it expands, this potential outcome can be simplified to $Y_{2016}^{\overline{Z}_{2014} = \mathbf{0}, Z_{2015} = 1}$. Note that in either case, we only need to define the vector of potential expansion statuses up to the year of expansion $t = g$. We can simplify notation and define the potential outcome in general terms as $$Y_t^{z;g}  =  Y_t^{\overline{Z}_{g-1} = \mathbf{0}, Z_g = z}.$$ If we let $z = 0$, we are considering the potential outcome under the hypothetical scenario of a state not yet expanding by year $g$ (note that this does not imply a state will never expand). If $z = 1$, we are considering the potential outcome under the hypothetical scenario of a state first expanding during year $g$.\\

\noindent  As an example, $Y_{2016}^{1;2016}$ represents the potential outcome in 2016 under the hypothetical scenario in which a state first expanded in 2016 and $Y_{2016}^{0;2016}$ represents the potential outcome in 2016 under the hypothetical scenario in which a state has not yet expanded Medicaid by 2016. We can then define the individual-level causal effect of first expanding Medicaid in 2016 on an individual in 2016 as $Y_{2016}^{1;2016} - Y_{2016}^{0;2016}$. More generally, the casual effect at time $t$ of first expanding Medicaid at time $g$ on an individual is $Y_t^{1;g} - Y_t^{0;g}$. Unfortunately, both $Y_t^{1;g}$ and $Y_t^{0;g}$ cannot be directly observed, given that a state will strictly have expanded Medicaid or not expanded Medicaid during any given year. The unobservable counterfactual has been referred to as the fundamental problem of causal inference.\cite{Holland} The average treatment effect (ATE) across all individuals is then $ATE(G = g, T = t) = E[Y_t^{1;g} - Y_t^{0;g}]$, interpreted as the average treatment effect at time $t$ of first expanding Medicaid at time $g$ among all individuals. We can also define the average treatment effect among the treated (ATT) as $ATT(G = g, T = t) = E[Y_t^{1;g} - Y_t^{0;g} | G = g]$, which can be interpreted as the average treatment effect at time $t$ of first expanding Medicaid at time $g$ among individuals living in states in expansion group $G = g$. We simplify the notation of $ATT(G = g, T = t)$ to $ATT(g,t)$ moving forward. Though both $Y_t^{1;g}$ and $Y_t^{0;g}$ cannot be directly observed, the $ATT(g,t)$ is identifiable under assumptions we detail in later sections.

\section{Two-Groups, Two-Periods}\label{2x2}
	\noindent We will first focus on the simple example in which we wish to estimate $ATT(G=2014,T=2014)$, or more concretely $$ATT(2014,2014) = E[Y_{2014}^{1;2014} - Y_{2014}^{0;2014} | G = 2014],$$ the average treatment effect in 2014 among those in states that first expanded in 2014.
 
\subsection{Difference-in-differences}\label{2x2 did}
	
 \noindent To identify $ATT(2014,2014)$ using DID techniques, we only need to consider $T = 2013, 2014$ and $G = 2014$, $>$2014, meaning we are in a two-period, two-group setting in which one group of states first expanded in 2014 and one group has not yet expanded in 2014.\\

\noindent Using DID to estimate $ATT(2014,2014)$ with observable data requires the consistency\cite{cole_frangakis_2009} and parallel trends assumptions. Specifically, in our two-group, two-period context, we assume

\begin{flalign*}
 &\mbox{A1. \textbf{Consistency}: }  Y_{2014}^{\overline{Z}_{2014} = \mathbf{z}} = Y_{2014} \mbox{ if } \overline{Z}_{2014} = \mathbf{z}&
\end{flalign*}

\noindent and 

\begin{flalign*}
 &\mbox{A2. \textbf{Unconditional Parallel Trends}: }  E[Y_{2014}^{0;2014} - Y_{2013}^{0;2014} | G = 2014] = E[Y_{2014}^{0;2014} - Y_{2013}^{0;2014} | G > 2014].&
\end{flalign*}
 
 \noindent Assumption A1 states that the potential outcome of an individual under their observed expansion status is the observed outcome for that individual. Assumption A2 states that the change over time in mean retention of those in an expansion state in the absence of expansion would be the same as the change over time in mean retention for those in non-expansion states. It does not imply that retention would be the same in both the post-expansion period (2014) and the pre-expansion period (2013) in the absence of expansion.\\

\noindent The parallel trends assumption is not directly testable in two-group, two-time period DID approaches. Therefore it is important to think conceptually about why it may or may not be valid within the context of the study. When there are more groups in more time periods, as in Section \ref{more groups and time}, the parallel trends assumption can be partially tested, although with limitations\cite{roth2022}, by testing the parallelism of pre-treatment trends.\cite{Wing} Note that the trends of either group do not have to be linear for the assumption to hold.\\

\noindent With assumptions A1 and A2, we can identify $ATT(2014,2014)$ with observable data as follows:

\begin{align*}
ATT(2014,2014) & = E[Y_{2014}^{1;2014}| G=2014] - E[Y_{2014}^{0;2014}  | G =2014]\\
&= E[Y_{2014}| G=2014] - E[Y_{2014}^{0;2014} | G =2014]\\
&= E[Y_{2014} | G = 2014] - E[Y_{2013}^{0;2014} |G=2014] - E[Y_{2014}^{0;2014} | G =2014] + E[Y_{2013}^{0;2014} |G=2014]\\
&= E[Y_{2014} | G = 2014] - E[Y_{2013}|G=2014] - E[Y_{2014}^{0;2014}  | G =2014] + E[Y_{2013}^{0;2014} | G=2014]\\
&= E[Y_{2014} - Y_{2013}| G = 2014] - E[Y_{2014}^{0;2014}  - Y_{2013}^{0;2014} | G =2014]\\
&= E[Y_{2014} - Y_{2013}| G = 2014] - E[Y_{2014}^{0;2014}  - Y_{2013}^{0;2014} | G > 2014]\\
&= E[Y_{2014} - Y_{2013}| G= 2014] - E[Y_{2014} - Y_{2013}| G > 2014]
\end{align*}

\noindent where lines 2, 4, and 7 are due to assumption A1 and line 6 is due to assumption A2.\\
	
\noindent  Given our sample, we estimate the $ATT(2014,2014)$ as $\widehat{ATT}(2014,2014) = \widehat{E}[Y_{2014} - Y_{2013} | G = 2014]- \widehat{E}[Y_{2014} - Y_{2013} | G > 2014] = \left( 0.729 - 0.736 \right) - \left( 0.753- 0.737\right) = -0.023$. This unadjusted estimate indicates expansion in 2014 resulted in a 2.3\% absolute decrease in retention for PWH in expansion states, with a 95\% confidence interval (CI) of a $1.2$\% to $3.3$\% absolute decrease. This confidence interval was calculated using a simple bootstrap approach that appropriately accounts for the clustered nature of the data (i.e., that the same individual could contribute to the data in 2013 and 2014). For the sake of brevity and presentation flow, we do not go into further detail about variance estimation here. Code for all analyses in this paper can be found at \url{https://biostat.app.vumc.org/wiki/Main/ArchivedAnalyses}.

\subsection{Linear Regression Model}\label{2x2 regression model}

\noindent Another approach to estimate $ATT(2014,2014)$ is through linear regression modeling. Consider the regression model below:

	\begin{equation}\label{2x2 regression model formula}
E[\mbox{Y}_{t}|G, T] = \beta_0 +  \beta_1 \textbf{I}\{G = 2014\} + \beta_2 \textbf{I}\{T = 2014\}+ \beta_3\textbf{I}\{G = 2014\} \times \textbf{I}\{T = 2014\}
	\end{equation}
	
\noindent such that the $ATT(2014,2014) = \beta_3$ under assumptions A1 and A2, which can be shown as follows:

	\begin{align*}
\beta_3 &= \left(\left(\beta_0 + \beta_1 + \beta_2 + \beta_3\right) - \left(\beta_0 + \beta_1 \right)\right) - \left(\left(\beta_0 + \beta_2\right) - \beta_0 \right)\\
&=  \left(E[Y_{2014}| G = 2014] -E[Y_{2013} | G = 2014]\right) - \left(E[Y_{2014}| G > 2014] -E[ Y_{2013} | G > 2014]\right)\\
&=  E[Y_{2014} - Y_{2013} | G = 2014] - E[Y_{2014} - Y_{2013} | G > 2014]\\
&= E[Y_{2014}^{1;2014} - Y_{2014}^{0;2014} | G = 2014] \mbox{ as shown in Section \ref{2x2 did}}\\
&= ATT(2014,2014).
\end{align*} 

\noindent   Our fitted model resulted in $\widehat{\beta}_3 = -0.023$ (95\% CI of $-0.034$ to $-0.012$). The point estimate obtained using model (\ref{2x2 regression model formula}) is equivalent to the point estimate obtained using DID as presented in Section \ref{2x2 did}, confirming the equivalence in this simple, unadjusted setting. While we used a bootstrap approach to calculate the confidence interval in Section \ref{2x2 did}, in this section, we calculated the confidence interval using an analytic approximation with robust standard errors that accounts for the clustered nature of the data. This was done by fitting model (\ref{2x2 regression model formula}) using the \texttt{ols} function and calculating robust standard errors using the \texttt{robcov} function from the \texttt{rms} package \cite{rms} in \texttt{R}. \cite{R} Though different, both approaches for calculating the 95\% CI are valid.

\section{Covariate adjustment in the two-group, two-period setting}\label{2x2 covariates}
\subsection{Difference-in-differences}\label{2x2 covariates did}

\noindent Suppose the unconditional parallel trends assumption (A2) is unreasonable such that the ATT cannot be identified using the method described in Section \ref{2x2}. Violation of the unconditional parallel trends assumption can be tied to differences in covariate distributions across expansion groups and covariate-specific trends in retention among expansion groups. In other words, we might believe that in the absence of expansion, retention trends would not be parallel between expansion and non-expansion groups due to covariate imbalances across groups. If this is the case, a more realistic parallel trends assumption conditional on covariates can be made to estimate the ATT.\cite{Heckman}\cite{Abadie}\cite{SANTANNA2020} A conditional parallel trends assumption allows for these covariate-specific trends in retention across expansion groups and is therefore a weaker assumption than the unconditional parallel trends. Let $\boldsymbol{X}$ represent a covariate or vector of covariates that is unaffected by expansion (e.g. race, age at enrollment, region, etc.). We assume 

\begin{flalign*}
 &\mbox{A3. \textbf{Conditional Parallel Trends}: }  E[Y_{2014}^{0;2014} - Y_{2013}^{0;2014} | \boldsymbol{X} = \boldsymbol{x}, G = 2014] = E[Y_{2014}^{0;2014} - Y_{2013}^{0;2014} | \boldsymbol{X} = \boldsymbol{x}, G > 2014] \mbox{ for all } \boldsymbol{x}.&
\end{flalign*}

\noindent Assumption A3 states that for individuals with the same observed covariate values in $\boldsymbol{X}$, the change over time in mean retention of those in an expansion state in the absence of expansion would be the same as the change over time in mean retention of those in non-expansion states. To non-parametrically identify the ATT in this context, in addition to Assumption A3, we make a common support assumption: that there is a non-zero probability of expanding and not expanding for all levels of covariate X. Specifically,

\begin{flalign*}
 &\mbox{A4. \textbf{Common Support}: } P(G>2014| \boldsymbol{X}=\boldsymbol{x}) > 0 \mbox{ and } P(G = 2014| \boldsymbol{X}=\boldsymbol{x}) > 0 \mbox{ for all }\boldsymbol{X}\mbox{ where } P(\boldsymbol{X}=\boldsymbol{x}) > 0.&
\end{flalign*}

 \noindent With assumptions A3 and A4, one can identify $ATT(2014,2014)$ incorporating covariates as follows \cite{Callaway}:

		\begin{align*}
	&ATT(2014,2014) \\
	&=  E[Y_{2014}^{1;2014} - Y_{2014}^{0;2014} | G = 2014]\\
	&= E[Y_{2014} - Y_{2013}| G = 2014] - E[Y_{2014}^{0;2014}  - Y_{2013}^{0;2014} | G =2014]\\
	&= E[Y_{2014} - Y_{2013}| G = 2014] - E\left[E[Y_{2014}^{0;2014}  - Y_{2013}^{0;2014} | \boldsymbol{X}=\boldsymbol{x}, G =2014] | G = 2014\right]  \\
	&= E[Y_{2014} - Y_{2013}| G = 2014] - E\left[E[Y_{2014}^{0;2014}  - Y_{2013}^{0;2014} | \boldsymbol{X}=\boldsymbol{x}, G >2014] | G = 2014\right] \\
	&= E[Y_{2014} - Y_{2013}| G = 2014] - E\left[E[Y_{2014}  - Y_{2013}| \boldsymbol{X}=\boldsymbol{x}, G > 2014] | G = 2014\right] \\
	&= E[Y_{2014} - Y_{2013}| G = 2014] - \left(E\left[E[Y_{2014} | \boldsymbol{X}=\boldsymbol{x}, G > 2014]  - E[Y_{2013}| \boldsymbol{X}=\boldsymbol{x}, G > 2014] | G = 2014\right]\right).
	\end{align*}

	\noindent The $ATT(2014,2014)$ is now written in terms of observable data. However, to estimate $E[Y_t | \boldsymbol{X}=\boldsymbol{x}, G > 2014, T = t]$ we may want to use a model, particularly if $X$ is multidimensional or continuous. One such linear model is $E[Y_{t} | \boldsymbol{X}=\boldsymbol{x}, G > 2014, T = t] = \beta_{0,t} + \beta_{1,t}X$. Because one model is fit for each year, this approach allows for covariate-specific trends over time since each covariate is implicitly interacted with time. One may then estimate $\beta_{0,t}$ and $\beta_{1,t}$ using data from individuals in non-expansion states and plug them in to estimate $ATT(2014,2014)$ as 

			\begin{align*}
	 &\widehat{ATT}(2014,2014) \\
	 &= \widehat{E}[Y_{2014} - Y_{2013}| G = 2014] - \left(\widehat{E}\left[\widehat{E}[Y_{2014} | \boldsymbol{X}=\boldsymbol{x}, G > 2014]  -\widehat{E}[Y_{2013}| \boldsymbol{X}=\boldsymbol{x}, G > 2014] | G = 2014]\right]\right)\\
	 &=  \widehat{E}[Y_{2014} - Y_{2013}| G = 2014] - \left(\frac{1}{\sum_{i = 1}^N I(G_i = 2014)}\sum_{i = 1}^{N}I(G_i = 2014)\left(\left(\widehat{\beta}_{0,2014} + \widehat{\beta}_{1,2014}x_i\right)  -\left(\widehat{\beta}_{0,2013} + \widehat{\beta}_{1,2013}x_i\right)\right)\right)
	\end{align*}

	\noindent where $N$ is the total number of observations and $I(G_i = 2014)$ is an indicator for being in the $G = 2014$ expansion group. For example, $\widehat{\beta}_{0,2014} + \widehat{\beta}_{1,2014}x_i$ represents the estimated value for retention in 2014 for an individual with covariates $x_i$ had they been in the non-expansion group. The summation standardizes this expectation across the covariate distribution of those in states that expanded. Thus, the final term of $\widehat{ATT}(2014,2014)$ is an estimate of the difference between mean predicted retention in 2014 and 2013 of those in expansion states had they not expanded.\\

	\noindent To obtain an unbiased estimate of the ATT, the model for $E\left[Y_{t} | \boldsymbol{X}=\boldsymbol{x}, G > 2014, T\right]$ must be properly specified. Less structured models for estimating $E\left[Y_{t} | \boldsymbol{X}=\boldsymbol{x}, G > 2014 , T\right]$ can also be fit, although fitting more flexible models requires more data. Other approaches based on inverse probability weights, including a doubly robust estimator,\cite{Callaway} exist to incorporate covariates into DID analyses, but for simplicity, this manuscript only considers the outcome regression approach. Incorporating age at enrollment, sex, race, and region through this approach results in an $\widehat{ATT}(2014,2014)$ of $-0.13$   (95\% CI of $-0.29$ to $0.032$).   Note that although covariates are incorporated in this estimate, the target parameter is the marginal ATT (i.e., not conditional on covariates).   Confidence intervals for ATT estimates are calculated using a clustered bootstrapping algorithm, detailed in Callaway and Sant'Anna 2021.   This approach, as well as the approach discussed in future Section \ref{more groups and time did}, can be implemented using the \texttt{att\_gt} function from the \texttt{did} package\cite{did_package} in \texttt{R}.\cite{R}

\subsection{Linear Regression Model}\label{2x2 covariates regression model}

\noindent Adjusting for covariates in the linear regression modeling setting is straightforward by simply including the covariates in model (\ref{2x2 regression model formula}); however, additional assumptions are necessary to identify and estimate the ATT from a single model coefficient. We first begin by adjusting for only one time-fixed variable, race, in the form of an indicator for being Black or non-Black. Our model of interest is then

	\begin{align}\label{2x2 covariates regression model formula}
E[\mbox{Y}_{t}| G, T, race] &= \beta_0^* +  \beta_1^* \textbf{I}\{G = 2014\} + \beta_2^* \textbf{I}\{T = 2014\}+ \beta_3^*\textbf{I}\{G = 2014\} \times \textbf{I}\{T = 2014\} + \beta_4^*\textbf{I}\{\text{Black}\}.
\end{align}

\noindent Before identifying the ATT, let's first consider two quantities to examine how race affects retention over time: the mean difference in retention between 2014 and 2013 among expansion states conditional on race and not conditional on race. It is easily shown that $E[Y_{2014} - Y_{2013} | G = 2014,\text{Black}]= E[Y_{2014} - Y_{2013} | G = 2014,\text{non-Black}]=E[Y_{2014} - Y_{2013} | G = 2014]= \beta_2^* + \beta_3^*$ under model (\ref{2x2 covariates regression model formula}). Under model (\ref{2x2 regression model formula}), which does not include race, the unconditional mean difference is $E[Y_{2014} - Y_{2013} | G = 2014]= \beta_2 + \beta_3$. Note that race does not affect the mean difference in retention over time in expansion states even when it is included in the model, $E[Y_{2014} - Y_{2013} | G = 2014]= \beta_2 + \beta_3 = \beta_2^* + \beta_3^*$. This can be verified empirically: $\hat \beta_2 + \hat \beta_3 = -0.007$ from model (\ref{2x2 regression model formula}) and $\hat \beta_2^* + \hat \beta_3^* = -0.007$ from model (\ref{2x2 covariates regression model formula}). Similarly, $E[Y_{2014} - Y_{2013} | G > 2014,\text{Black}] = E[Y_{2014} - Y_{2013} | G > 2014,\text{non-Black}] = E[Y_{2014} - Y_{2013} | G > 2014]$. Therefore, including time-invariant covariates into model (\ref{2x2 regression model formula}), as in model (\ref{2x2 covariates regression model formula}), does not allow for those covariates to affect outcome trends over time.\\

\noindent Recall assumption A2 (unconditional parallel trends), which makes assumptions about unconditional trends of the outcome over time. Because model (\ref{2x2 covariates regression model formula}) does not allow for trends of the outcome over time to be affected by race, assumption A2 must still hold when identifying the ATT in this setting. Thus, under assumptions A1 and A2, we can identify $ATT(2014,2014)$.\\

\begin{align*}
ATT(2014,2014) &=E[Y_{2014}^{1;2014} - Y_{2014}^{0;2014} | G = 2014]\\
&=E\left[E[Y_{2014}^{1;2014} - Y_{2014}^{0;2014} | G = 2014, race]| G = 2014\right]\\
&=E\left[E[Y_{2014} - Y_{2013} | G = 2014, race] - E[Y_{2014}^{0;2014} - Y_{2013}^{0;2014} | G = 2014, race]| G = 2014\right]\\
&=E\left[E[Y_{2014} - Y_{2013} | G = 2014] - E[Y_{2014}^{0;2014} - Y_{2013}^{0;2014} | G = 2014]| G = 2014\right]\\
&=E\left[E[Y_{2014} - Y_{2013} | G = 2014] - E[Y_{2014}^{0;2014} - Y_{2013}^{0;2014} | G > 2014]| G = 2014\right]\\
&=E\left[E[Y_{2014} - Y_{2013} | G = 2014] - E[Y_{2014} - Y_{2013} | G > 2014]| G = 2014\right]\\
&=E[\beta_3^* | G = 2014]\\
&= \beta_3^*
\end{align*} 

\noindent Again, because race does not affect the change in retention over time in this model setup, $\beta_3^*$ from model (\ref{2x2 covariates regression model formula}) is equivalent to $\beta_3$ from model (\ref{2x2 regression model formula}).   Both models result in $\widehat{\beta}_3^*= \widehat{\beta}_3 = -0.023$ (95\% CI of $-0.034$ to $-0.012$), meaning the inclusion of race does not affect the estimate of the ATT (confidence intervals in this section are calculated using the same approach described in Section \ref{2x2 regression model}).   In other words, this modeling approach forces the treatment effect to be the same across all values of race. Therefore, in this setting, we are implicitly assuming

\begin{flalign*}
 &\mbox{A5. \textbf{Homogeneous Treatment Effects Across Covariates}: } ATT(g,t)\mbox{ is constant across all values of } \boldsymbol{X}.&
\end{flalign*}

\noindent In contrast, the approach discussed in Section \ref{2x2 covariates did} does not assume the ATT is constant across all covariates, even though it also results in a marginal ATT. Instead, it averages over conditional ATT estimates to get a marginal ATT, which does not rely on assumption A5.\\

\noindent In our setting, race is defined as an indicator for being Black or not Black, therefore it is straightforward to assess the validity of assumption A5. We can fit the unadjusted model (\ref{2x2 regression model formula}) among Black individuals and among non-Black individuals separately and compare the estimated treatment effect of each group.   Doing this results in $\widehat{ATT}(2014,2014|\text{Black}) = -0.014$ (95\% CI of $-0.03$ to 0.002) and $\widehat{ATT}(2014,2014|\text{non-Black}) = -0.024$ (95\% CI of $-0.038$ to $-0.01$).   While these estimates are not identical, assumption A5 seems reasonable in this case. Note that assessing the validity of assumption A5 is not always possible if a covariate has many categories or is continuous.\\

\noindent Now consider a situation in which we want to allow race to affect the trends of the outcome over time. To allow for this, we can include an interaction between race and the indicator for being in the post-treatment time period into model (\ref{2x2 regression model formula}). Specifically, we include an interaction between the indicator for being Black or non-Black and the indicator for being in the year 2014 and thus allow trends in retention over time to differ by race. Our model is then

\begin{align}\label{model 4.2 2}
E[\mbox{Y}_{t}| G, T, race] &= \beta_0^{**} +  \beta_1^{**} \textbf{I}\{G = 2014\} + \beta_2^{**} \textbf{I}\{T = 2014\}+ \beta_3^{**}\textbf{I} \{G = 2014\} \times \textbf{I}\{T = 2014\} + \beta_4^{**}\textbf{I}\{\text{Black}\}\\
& + \beta_5^{**}\textbf{I}\{\text{Black}\}\times \textbf{I}\{T = 2014\}.  \notag
\end{align}

\noindent Once again consider the mean difference in retention between 2014 and 2013 among expansion states conditional on race and not conditional on race. Under model (\ref{model 4.2 2}), the mean differences conditional on race are $E[Y_{2014} - Y_{2013} | G = 2014, \text{Black}] = \beta_2^{**} + \beta_3^{**} +\beta_5^{**}$ and $E[Y_{2014} - Y_{2013} | G = 2014, \text{non-Black}] = \beta_2^{**} + \beta_3^{**}$, which now differ depending on the value of race. The mean difference unconditional on race is $E[Y_{2014} - Y_{2013} | G = 2014] = \beta_2^{**} + \beta_3^{**} +\beta_5^{**}E[\textbf{I}\{\text{Black}\}|G=2014]$. Therefore by construction, model (\ref{model 4.2 2}) now allows for the trends in the outcome over time to be affected by the covariate. Because of this, the parallel trends assumption no longer holds unconditionally and we must rely on assumption A3 (conditional parallel trends assumption) to account for the fact that race can now affect retention trends over time. Based on model (\ref{model 4.2 2}), we can identify $ATT(2014,2014)$ with assumptions A1 and A3.

\begin{align*}
ATT(2014,2014) &=E[Y_{2014}^{1;2014} - Y_{2014}^{0;2014} | G = 2014]\\
&=E\left[E[Y_{2014}^{1;2014} - Y_{2014}^{0;2014} | G = 2014, race]| G = 2014\right]\\
&=E\left[E[Y_{2014} - Y_{2013} | G = 2014, race] - E[Y_{2014}^{0;2014} - Y_{2013}^{0;2014} | G = 2014, race]| G = 2014\right]\\
&=E\left[E[Y_{2014} - Y_{2013} | G = 2014, race] - E[Y_{2014}^{0;2014} - Y_{2013}^{0;2014} | G > 2014,race]| G = 2014\right]\\
&=E\left[E[Y_{2014} - Y_{2013} | G = 2014, race] - E[Y_{2014} - Y_{2013} | G > 2014,race]| G = 2014\right]\\
&= E\left[\left((\beta_0^{**} + \beta_1^{**} + \beta_2^{**} + \beta_3^{**} + \beta_4^{**}\textbf{I}\{\text{Black}\} + \beta_5^{**}\textbf{I}\{\text{Black}\}) - (\beta_0^{**} + \beta_1^{**}   + \beta_4^{**}\textbf{I}\{\text{Black}\}) \right)\right.\\
&\hspace{0.5 cm}- \left.\left((\beta_0^{**} + \beta_2^{**} + \beta_4^{**}\textbf{I}\{\text{Black}\} + \beta_5^{**}\textbf{I}\{\text{Black}\}) - (\beta_0^{**} +  \beta_4^{**}\textbf{I}\{\text{Black}\}) \right)| G = 2014\right]\\
&= E[\beta_3^{**}| G = 2014]\\
&= \beta_3^{**}
\end{align*} 

\noindent Lines 3 and 5 hold from assumption A1 and line 4 holds from assumption A3. Fitting model (\ref{model 4.2 2}) results in the estimate $\widehat{\beta}_3^{**} = -0.019$   (95\% CI of $-0.03$ to $-0.008$).   This indicates that under the parallel trends assumption conditional on race, Medicaid expansion resulted in an estimated 1.9\% absolute decrease in retention for PWH in expansion states. Notice how we are still implicitly making the homogeneous treatment effect assumption (A5); the ATT is still assumed to be the same across both race groups, as all race group parameters in line 6 cancel out with each other. As discussed earlier, assumption A5 does not appear unreasonable after comparing the estimated ATT among both race groups.\\

\noindent Assumption A5 may not always be reasonable, however.   For example, fitting model (\ref{2x2 regression model formula}) within each region group results in conditional ATT estimates of 0.068 (95\% CI of $0.044$ to $0.092$), $-0.36$ (95\% CI of $-0.89$ to $0.18$),  $-0.023$ (95\% CI of $-0.045$ to $-0.001$), and $-0.11$ (95\% CI of $-0.41$ to $0.20$) for those among the SO region, WE region, NE region, and MW region, respectively.   In this case, it seems unreasonable to assume the ATT is constant across all four region groups. Therefore, fitting model (\ref{model 4.2 2}) with region as the covariate instead of race would likely result in a biased estimate of the overall ATT, even if we believe the conditional parallel trends assumption conditional on region holds.\cite{SANTANNA2020}\cite{ROTH2023} If assumption A5 is thought to be violated, the linear modeling approach presented in this section is likely not the best approach to use, given it may provide biased estimates of the ATT. Other approaches exist, such as the approach described in Section \ref{2x2 covariates did}, that rely on the conditional parallel trends assumption without making any homogeneity assumptions. \cite{SANTANNA2020}\cite{ROTH2023}\\

\noindent \textbf{Table \ref{tab:Table 4.2}} displays $ATT(2014,2014)$ estimates and 95\% confidence intervals from the linear regression modeling approach described in Section \ref{2x2 covariates regression model} and the DID approach described in Section \ref{2x2 covariates did}. To summarize, regardless of the nature of the covariates being included or the type of parallel trends assumption being made in the linear model set-up, unbiased estimation of the ATT relies on the assumption that the ATT does not vary across different values of the covariates. As noted, the DID approach does not rely on this homogeneity assumption, but rather averages over conditional ATT estimates.   For example, when region is included as a covariate in the DID approach, the confidence interval becomes much wider, likely due to the fact that ATT estimates vary across regions.   The DID approach in Section \ref{2x2 covariates did} adjusts for covariates only within the control group and then standardizes across covariates in the treatment group while implicitly interacting all covariates with time. The linear model approach adjusts for covariates within both the control group and the treatment group in one model and does not implicitly interact each covariate with time unless specified. While the DID approach relies on the conditional parallel trends assumption, the linear modeling approach relies on the unconditional parallel trends assumption, unless a covariate is interacted with time in the model. These differences between the approaches result in different estimates of the ATT. Regardless of these differences, both approaches can produce a valid estimate of $ATT(2014,2014)$ if researchers understand and believe the assumptions being made.\\

\begin{table}[ht]
\centering
\caption{  Estimates and 95\% confidence intervals of $ATT(2014,2014)$ using the linear regression modeling approach described in Section \ref{2x2 covariates regression model} and the DID approach described in Section \ref{2x2 covariates did}. Details of the identification of $ATT(2014,2014)$ for the model including all covariates interacted with time (last row) are in the Appendix in Section \ref{app table1}. Covariates considered, as noted, include age at enrollment (centered and  modeled using splines with 3 knots), sex (male or female), race (non-Black or Black), region (WE, NE, SO, or MW). }
\begin{tabular}{|l|cc|}
\hline
                                                                                                                     & \multicolumn{2}{c|}{\textbf{$\widehat{ATT}(2014,2014)$}}                                  \\ \hline
\textbf{Covariates in Model}                                                                                         & \multicolumn{1}{c|}{\textbf{Linear Regression Model}} & \textbf{Difference-in-differences}            \\ \hline
Unadjusted                                                                                                                 & \multicolumn{1}{c|}{$-0.023$ ($-0.03,-0.01$)}              & {$-0.023$ ($-0.03,-0.01$)} \\ \hline
Race                                                                                                                 & \multicolumn{1}{c|}{$-0.023$ ($-0.03,-0.01$)}              & \multirow{2}{*}{$-0.020$ ($-0.03,-0.01$)} \\ \cline{1-2}
Race, Race $\times$ Time                                                                                                    & \multicolumn{1}{c|}{$-0.019$ ($-0.03,-0.01$)}              &                         \\ \hline
Region                                                                                                                  & \multicolumn{1}{c|}{$-0.023$ ($-0.03,-0.01$)}              & \multirow{2}{*}{$-0.14$ ($-0.30,0.03$)}  \\ \cline{1-2}
Region, Region $\times$ Time                                                                                                      & \multicolumn{1}{c|}{$0.037$ ($0.02$ , $0.05$)}               &                         \\ \hline
Age, Sex, Race, Region                                                                                               & \multicolumn{1}{c|}{$-0.023$ ($-0.03,-0.01$)}              & \multirow{2}{*}{$-0.13$ ($-0.30,0.04$)} \\ \cline{1-2}
\begin{tabular}[c]{@{}l@{}}Age, Age $\times$ Time, Sex, Sex $\times$ Time,\\ Race, Race $\times$ Time, Region, Region $\times$ Time\end{tabular} & \multicolumn{1}{c|}{$0.037$ ($0.02$,$0.05$)}               &                         \\ \hline
\end{tabular}
\label{tab:Table 4.2}
\end{table}

\section{Generalizations to more groups and time periods}\label{more groups and time}

\noindent Sections \ref{2x2} and \ref{2x2 covariates} detail the estimation of the ATT with only two groups and two time periods. In the case of Medicaid expansion, there are more than two expansion groups and two time periods; states expanded Medicaid in a staggered format from 2014 to 2016 in this data set. We are interested in summarizing the overall effect of expansion across all groups and time periods. 

\subsection{Difference-in-differences}\label{more groups and time did}

\noindent Until now, we have focused on calculating the effect of expansion in the first year of expansion. However, the methods described in Sections \ref{2x2 did} and \ref{2x2 covariates did} can be used to calculate the effect of expansion during any pre-expansion and post-expansion year. At the end of this section, we detail how the ability to calculate the effect of expansion during pre-expansion periods can be used as a proxy for testing the parallel trends assumption. For now, we focus on calculating effects during post-expansion periods. For example, we can estimate $ATT(2014,2015), ATT(2014,2016),$ and $ATT(2014,2017)$, the effects of expansion in 2014 during the years 2015, 2016, and 2017, respectively. In other words, we can calculate the effect of expansion in 2014 after any number of years since expansion. Let $w \in \{0,1,2,3\}$ represent the number of years since expansion. With this, $ATT(2014,2017)$ can be written as $ATT(2014, 2014 + 3)$ and so $w = 3$ in this case. This concept can be applied to any expansion group and time since expansion, so we can write the effect of expansion more generally as $ATT(g, g + w) = E[Y_{g+w}^{1;g} - Y_{g+w}^{0;g} | G = g]$ for $g+w \leq 2017$. Limiting to $g + w \leq 2017$ ensures we only consider ATT estimates within the time period of the study. If we wish to include covariates, we can use the approach described in  Section \ref{2x2 covariates did} to identify each $ATT(g,g+w)$ as 
\begin{align*}
ATT(g, g + w)&= E[Y_{g+w} - Y_{g-1}| G = g] - \left(E\left[E[Y_{g+w} | \boldsymbol{X}=\boldsymbol{x}, G > g\right]  - E\left[Y_{g-1}| \boldsymbol{X}=\boldsymbol{x}, G > g] | G = g\right]\right).
\end{align*}

\noindent As described in Section \ref{2x2 covariates did}, assumptions A1 and A3 must hold. In this context, consistency assumes that $Y_{g+w}^{\overline{Z}_{g+w} = \mathbf{z}} = Y_{g+w}$ if $\overline{Z}_{g+w} = \mathbf{z}$ and the conditional parallel trends assumption is $E[Y_{g+w}^{0;g} - Y_{g-1}^{0;g} | G = g, \mathbf{X}] = E[Y_{g+w}^{0;g} - Y_{g-1}^{0;g} |  G > g, \mathbf{X}]$ for $g + w \leq 2017$. Note that to identify $ATT(g, g+w)$ for all $g + w \leq 2017$, these two assumptions must hold for multiple groups.\\

\noindent There are many approaches available to calculate an aggregate version of the ATT that group expansion effects either with respect to time since expansion or expansion time.\cite{Callaway} In this section we detail an approach in Callaway \& Sant'Anna 2021 that focuses on grouping expansion effects with respect to $w$ to create a weighted average of ATT estimates calculated at each year since expansion. The first step is to calculate a collection of $ATT(g,g+w)$ estimates for all possible expansion year and time combinations.\\

\noindent Each $ATT(g, g + w)$ estimate can then be grouped with respect to $w$. We may then be interested in the average effect of expansion for each year since expansion $w$, specifically the expectation of all corresponding ATT values with respect to $g$: $$ATT_{g}(w) = E\left[ATT(g, g + w)\right],$$ which can be estimated using observable data as $$\widehat{ATT_{g}}(w) = \sum_{g}\hat{P}(G = g|G \leq 2017) \widehat{ATT}(g,g + w).$$ For example, including covariates race, age, sex, and region, the average effect of expansion during the year of expansion ($w = 0$) is calculated as 
\begin{align*}
\widehat{ATT_{g}}(0) &=  \widehat{P}(G = 2014|G \leq 2017)\widehat{ATT}(2014, 2014) +  \widehat{P}(G = 2015| G\leq 2017)\widehat{ATT}(2015, 2015)\\
&\hspace{0.5 cm}+  \widehat{P}(G = 2016|G \leq 2017)\widehat{ATT}(2016, 2016)\\
&= (0.91)(-0.13) + (0.08)(-0.20) + (0.01)(0.22)\\
&= -0.13.
\end{align*}

\noindent  The 95\% CI for $\widehat{ATT_{g}}(0)$ is $-0.28$ to $0.01$, once again calculated using a bootstrap approach detailed in Callaway and Sant'Anna 2021. Using the same process, the average effect of expansion one, two, and three years after expansion is $\widehat{ATT_{g}}(1) = -0.15$ (95\% CI of $-0.39$ to $0.09$), $\widehat{ATT_{g}}(2) = -0.22$ (95\% CI of $-0.48$ to $0.037$), and $\widehat{ATT_{g}}(3) = -0.27$ (95\% CI of $-0.53$ to $-0.007$), respectively. Given these results, it appears that the effect of expansion on retention becomes stronger with time.  \\

\begin{figure}[ht]
			\centering
			\includegraphics[width = 11 cm]{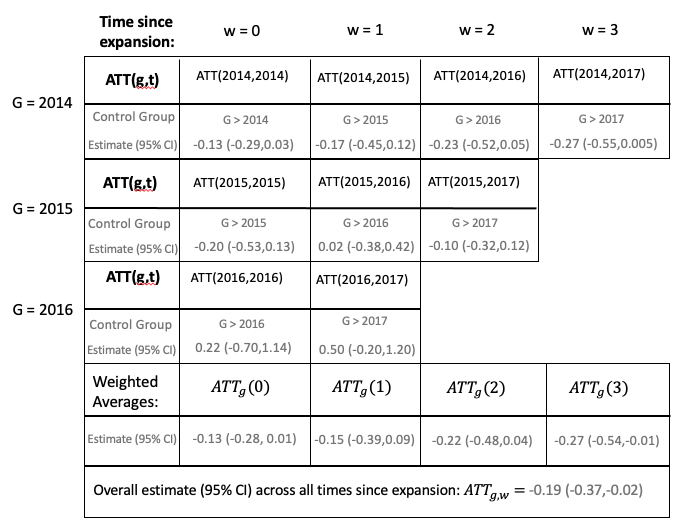}
			\caption{Set-up of aggregation approach, including estimate of each $\widehat{ATT}(g, g+ w)$ adjusting for age at enrollment (centered and  modeled using splines with 3 knots), sex (male or female), race (non-Black or Black), region (WE, NE, SO, or MW).}
   \label{fig:more time and groups did figure}
		\end{figure}

\noindent We end up with four $\widehat{ATT_{g}}(w)$ values because there are four possible values of $w$. One may then be interested in summarizing the ATT across all $w$. One summary is the simple average across the four $ATT_{g}(w)$ as $$ATT_{g,w} = \frac{1}{4}\sum_{w}ATT_{g}(w).$$  The estimation of $ATT_{g,w}$ takes into account the staggered adoption of Medicaid expansion and is a measure of the overall effect of expansion on clinical retention that uses information from all expansion groups and all post-expansion periods. Note that it assigns equal importance to each year since expansion, $w$, and hence may or may not be relevant or an oversimplification in some applications. $ATT_{g,w}$ can be estimated as $\widehat{ATT}_{g,w} = \frac{1}{4}\sum_{w}\widehat{ATT_{g}}(w) = \frac{1}{4}(-0.19 + -0.14 + -0.22 + -0.27) = -0.19$   (95\% CI of $-0.37$ to $-0.017$). This indicates an estimated overall negative effect of expansion on clinical retention among PWH.  \textbf{Figure \ref{fig:more time and groups did figure}} provides a visual representation of the approach and includes each corresponding $ATT(g, g+ w)$ estimate.\\

\noindent The conditional parallel trends assumption is untestable, however, testing pre-expansion trends can be used as a proxy to assess its validity. One way to do so in the setting of multiple groups and time periods is to follow the same steps described above using only pre-expansion information to then compare against a null hypothesis of no significant effect at each time period before expansion. For example, we can estimate $ATT(2014,2012)$ and $ATT(2014,2013)$, the effects of expansion in 2014 during the years 2012 and 2013 and group them into $w = -2$ and $w = -1$, respectively. In the case of testing parallel trends, each $ATT(g, g + w)$ estimate is grouped with respect to $w$, however each $w$ is now negative because we are calculating ATT estimates during years before expansion. We test the parallelism of pre-expansion trends by comparing all $ATT_{g}(w)$ using pre-expansion information against the null $ATT_{g}(w) = 0$. Evidence that points to any $ATT_{g}(w) \ne 0$ in this setting is an indication that trends in the outcome prior to expansion may not be parallel and therefore may not be parallel in the absence of expansion during post-expansion time periods. In our case, there was no indication of a violation of the conditional parallel trends assumption (p = 0.91). More details on pre-expansion trends are in Section \ref{parallel trends test} of the Appendix.

\subsection{Linear Regression Model}\label{more groups and time regression model}

We now focus on the two-way fixed effects (TWFE) model, a linear regression approach commonly used to estimate the ATT when there are more than two time periods and two treatment groups \cite{Wing}. It is a straightforward extension of model (\ref{2x2 regression model formula}) from Section \ref{2x2 regression model}. We will discuss the major assumption needed for this approach to result in a valid estimate of the ATT and the implications of this assumption not being met.\\

\noindent TWFE models include the time-varying indicator for treatment, group-fixed terms, and time-fixed terms. In the two-group, two-period setting described in Sections \ref{2x2 regression model} and \ref{2x2 covariates regression model}, the time-varying indicator for being in the 2014 expansion group during the post-expansion year 2014 is represented by $\textbf{I}\{G = 2014\} \times \textbf{I}\{T = 2014\}$. In the case of more than two groups and two time periods, we must define a more general time-varying indicator variable for being in an expansion group during a post-expansion period. We define this as $\textbf{I}\{expanded_t\} = \textbf{I}\{t \geq g\}$. For those in $G = 2015$, for example, $\textbf{I}\{expanded_{2014}\} = 0$, $\textbf{I}\{expanded_{2015}\} = 1$, and $\textbf{I}\{expanded_{2016}\} = 1$. For those in states that never expand, $\textbf{I}\{expanded_{t}\} = 0$ for all $t$. In our unadjusted three-group, three-period setting, the TWFE model is

\begin{align}\label{more groups more time regression model}
E[Y_t| G, T] = \beta_0' + \beta_1'\textbf{I}\{expanded_t\} + \beta_2'\textbf{I}(G = 2014) + \beta_3'\textbf{I}(T = 2014) + \beta_4'\textbf{I}(G = 2015) + \beta_5'\textbf{I}(T = 2015)
\end{align}

\noindent where $\beta_1'$ is said to be a treatment effect \cite{Wing}. To understand what $\beta_1'$ is actually estimating, we consider $ATT(2014,2014)$,  $ATT(2014,2015)$, and $ATT(2015,2015)$. Under assumptions A1 and A2, we have

\begin{align*}
ATT(2014,2014) &= E[Y_{2014} - Y_{2013} | G = 2014] - E[Y_{2014} - Y_{2013} | G > 2014]\\
&=\left((\beta_0' + \beta_1' + \beta_2' + \beta_3') - (\beta_0' + \beta_2')\right) - \left((\beta_0' + \beta_3') -\beta_0'\right)\\
&= \beta_1'
\end{align*}

\begin{align*}
ATT(2014,2015) &= E[Y_{2015} - Y_{2013} | G = 2014] - E[Y_{2015} - Y_{2013} | G > 2014]\\
&=\left((\beta_0' + \beta_1' + \beta_2' + \beta_5') - (\beta_0' + \beta_2')\right) - \left((\beta_0' + \beta_5') -\beta_0' \right)\\
&= \beta_1'
\end{align*}

\begin{align*}
ATT(2015,2015) &= E[Y_{2015} - Y_{2014} | G = 2015] - E[Y_{2015} - Y_{2014} | G > 2015]\\
&=\left((\beta_0' + \beta_1' + \beta_4' + \beta_5') - (\beta_0' + \beta_3' + \beta_4')\right) - \left((\beta_0' + \beta_5') -(\beta_0' + \beta_3') \right)\\
&= \beta_1'
\end{align*}

\noindent Notice how $ATT(2014,2014)$, $ATT(2014,2015)$, and $ATT(2015,2015)$ all equal $\beta_1'$. Thus, when we use the TWFE approach, we are assuming 

\begin{flalign*}
 &\mbox{A6. \textbf{Homogeneous Treatment Effects Across Treatment Groups and Time}: } ATT(g,t)\mbox{ is constant across all treatment}\\
 &\mbox{groups } G \mbox{ and periods } T.&
\end{flalign*}

\noindent In our context, assumption A6 states that the effect in 2014 and 2015 of expanding Medicaid in 2014 is equivalent to the effect in 2015 of expanding Medicaid in 2015. If we believe this assumption is reasonable, then the TWFE model gives us a valid single-coefficient estimate ($\beta_1'$) of the effect of Medicaid expansion on clinical retention. A confidence interval for $\widehat{\beta}_1'$ can be obtained as described in Section \ref{2x2 regression model}.\\

\noindent If the treatment effects across groups and time are not homogeneous, and therefore assumption A6 is violated, it can be shown that $\widehat{\beta}_1'$ is not an estimate of the ATT, but rather a weighted average of four different two-year, two-group comparisons. \cite{GoodmanBacon} Some comparisons are not meaningful because they use an already-treated group as the control group. In addition, the weights associated with each comparison are not easily interpretable and depend on the size of the overall sub-sample and relative sizes of the treatment and control groups used in each comparison, along with the timing of the treatment in each sub-sample.\cite{GoodmanBacon} Extensive work has been done to understand why the TWFE model is not robust to treatment effect heterogeneity. \cite{GoodmanBacon}\cite{deChaisemartin2020}\cite{Borusyak2022}\cite{Sun2021}\\

\noindent  The estimate of $\widehat{\beta}_1'=-0.028$   (95\% CI of $-0.037$ to $-0.019$) from model (\ref{more groups more time regression model}) can be interpreted as $\widehat{ATT}(2014,2014)$, $\widehat{ATT}(2014,2015)$, or $\widehat{ATT}(2015,2015)$ if the unconditional parallel trends assumption (A2) within each two-by-two comparison and the homogeneous treatment effect assumption (A6) hold.    In other words, $-0.028$ is the estimated ATT for those in states that expanded in either 2014 or 2015 during 2014-2015. If assumption A6 does not hold, the estimate of $-0.028$ is likely not an estimate of the overall ATT, but rather a combination of two-by-two comparisons without a clear causal interpretation. If a researcher has reason to believe the assumptions of unconditional parallel trends and homogeneous treatment effects hold in their context, then the unadjusted TWFE model is a straightforward method to estimate the overall ATT. Otherwise, the DID method described in Section \ref{more groups and time did} may be more appropriate. Note that the $ATT_{g,w}$ presented in Section \ref{more groups and time did} also combines $ATT$ across groups and years, which may not be desirable with heterogeneous effects. However, the manner in which $ATT_{g,w}$ combines $ATT$ is straightforward, transparent, and leads to an interpretable causal effect without making any homogeneity assumptions.

\section{Discussion}\label{conclusion}

\noindent In this paper, we detailed the identification and estimation of the ATT using a difference-in-differences approach and a linear regression modeling approach in the context of Medicaid expansion and clinical retention among PWH. We discussed the assumptions necessary for each approach to result in a valid, straightforward estimate of the ATT under increasingly complex scenarios involving the staggered nature of Medicaid expansion and the inclusion of covariates. Note that these estimates are presented here to illustrate the methods discussed in this paper. Further consideration of covariates and assumptions revolving around our research question would be necessary to make any definitive conclusions about the effect of Medicaid expansion on clinical retention among PWH in the United States.\\

\noindent We note that it may be tempting to include a time-varying covariate, such as CD4 count over time, into DID or linear regression approaches when such information is available. The appeal of doing so in the linear regression modeling approach would be to rely on the conditional parallel trends assumption. However, doing so in either approach comes with further assumptions and restrictions on the data. Finding an unbiased estimate of the ATT from one model parameter in the presence of a time-varying covariate requires additional assumptions about the path of the untreated potential outcome in relation to the covariate.\cite{caetano2022difference} In the case when a time-varying covariate is affected by the treatment, it becomes even more difficult to find an unbiased estimate of the ATT. \cite{caetano2022difference}\cite{Zeldow2019} Alternative, more complicated estimation approaches are available in the presence of such time-varying covariates that avoid these more restrictive assumptions.\cite{caetano2022difference}\\

\noindent A summary of the assumptions tied to each approach and its estimate of the ATT in 2014 is displayed in \textbf{Table \ref{tab:Table 5}}. Code for each approach, along with all analyses described in this paper, can be found at \url{https://biostat.app.vumc.org/wiki/Main/ArchivedAnalyses}. The approaches discussed in this paper are only two of the many possible causal inference approaches for observational data that can provide a valid policy effect estimate, and we focused only on the estimation of the ATT. The ATT is an estimate of the effect of expansion on clinical retention among the population of PWH in states that expanded Medicaid, therefore the population to which the ATT can be applied does not extend to PWH in non-expansion states. The ATE, in contrast, is an estimate of the effect of expansion on clinical retention among PWH living in any state, whether or not it expanded Medicaid, and would therefore be of interest to policymakers who wish to understand the impact Medicaid expansion would have on the population of PWH in any state. Another potential estimand is the average treatment effect among the untreated (ATU), which is an estimate of the effect of Medicaid expansion among the population of PWH in states that have not yet expanded Medicaid and is the most relevant estimand for policymakers in states that have not expanded Medicaid who are considering whether to expand Medicaid. The methods discussed in this paper provide ATT estimates, which may not always be the most relevant estimand for policymakers, therefore different methods that provide either an ATE or ATU estimate may be better suited. Greifer \& Stuart 2021 offer details on the difference in use and relevance of ATT, ATE, and ATU in the context of matching or weighting in observational studies. Haber et. al 2021 provide a summary of other policy impact evaluation designs and analyses in the context of the COVID-19 pandemic, including cross-sectional analyses, pre/post analyses, and interrupted time series. Approaches using propensity scores are also widely used to estimate policy effects with observational data.\cite{Austin2011}  Synthetic control approaches for staggered treatment adoption also exist.\cite{Ben-Michael} It is important to understand the relevance, limitations, and required assumptions of whichever approach and corresponding estimand is used to assess the causal impact of a policy. Not doing so may lead to invalid estimates and misguided policy-making decisions.

\begin{table}[ht]
	\def\arraystretch{1.5}
			\caption{List of approaches, covariate type, estimate of $ATT(2014,2014)$, and corresponding assumptions necessary for the valid estimation of $ATT(2014,2014)$ as listed. }

	\centering

	\begin{tabular}{|l|l|l|l|}
		\hline
		\textbf{Approach} & \textbf{\begin{tabular}[c]{@{}l@{}}Type of \\ Covariates\end{tabular}} & \textbf{$\mathbf{\widehat{ATT}(2014,2014)}$}                                                                                                                                                                                                             & \textbf{Assumptions Needed}                                                                                                                                                                                  \\ \hline \hline
		\begin{tabular}[c]{@{}l@{}} Simple DID \\ (Section \ref{2x2 did}) \end{tabular}   & None                                                                   & \begin{tabular}[c]{@{}l@{}}$\widehat{E}[Y_{2014} - Y_{2013}| G= 2014]$\\ $ - \widehat{E}[Y_{2014} - Y_{2013}| G > 2014]$\end{tabular}                                                                                                                            & A1, A2                                                                                                                                      \\ \hline
		\begin{tabular}[c]{@{}l@{}} Simple DID w/ \\ standardization\\ (Section \ref{2x2 covariates did}) \end{tabular}       & \begin{tabular}[c]{@{}l@{}}Time-fixed\end{tabular}      & \begin{tabular}[c]{@{}l@{}}$\widehat{E}[Y_{2014} - Y_{2013}| G = 2014]$\\ $ - \left(\widehat{E}\left[\widehat{E}[Y_{2014} | \boldsymbol{X}=\boldsymbol{x}, G > 2014\right]\right.$\\ $  - \left.\widehat{E}\left[Y_{2013}| \boldsymbol{X}=\boldsymbol{x}, G > 2014] | G = 2014\right]\right)$\end{tabular}         & A1, A3                                                                                                            \\ \hline
		\begin{tabular}[c]{@{}l@{}} Linear model \\ (Section \ref{2x2 regression model})   \end{tabular}     & None                                                                   & \begin{tabular}[c]{@{}l@{}}$\widehat{\beta}_3$ from $E[\mbox{Y}_{t}|G, T] = \beta_0 +  \beta_1 \textbf{I}\{G = 2014\} $\\ $+ \beta_2 \textbf{I}\{T = 2014\}$\\ $+ \beta_3(\textbf{I}\{G = 2014\} \times \textbf{I}\{T = 2014\}$\end{tabular}                      &A1, A2                                                                                                                                      \\ \hline
		\begin{tabular}[c]{@{}l@{}} Linear model  \\(Section \ref{2x2 covariates regression model})   \end{tabular}    & Time-fixed                                                             & \begin{tabular}[c]{@{}l@{}}$\widehat{\beta}_3^{*}$ from $E[\mbox{Y}_{t}|G, T, X] = \beta_0^{*} +  \beta_1^{*} \textbf{I}\{G = 2014\} $\\ $+ \beta_2^{*} \textbf{I}\{T = 2014\}$\\ $+ \beta_3^{*}(\textbf{I}\{G = 2014\} \times \textbf{I}\{T = 2014\} + \beta_4^{*} X$\end{tabular}   & A1, A2, A5                                                                                                            \\ \hline
		\begin{tabular}[c]{@{}l@{}} Linear model \\(Section \ref{2x2 covariates regression model})      \end{tabular}   & \begin{tabular}[c]{@{}l@{}}Time-fixed\\ interacted with time \end{tabular}                                                           & \begin{tabular}[c]{@{}l@{}}$\widehat{\beta}_3^{**}$ from $E[\mbox{Y}_{t}|G, T, X] = \beta_0^{**} +  \beta_1^{**} \textbf{I}\{G = 2014\} $\\ $+ \beta_2^{**} \textbf{I}\{T = 2014\}$\\ $+ \beta_3^{**}(\textbf{I}\{G = 2014\} \times \textbf{I}\{T = 2014\}+ \beta_4^{**} X$\\$ + \beta_5^{**} \boldsymbol{X}\times \textbf{I}\{T = 2014\}$\end{tabular} & A1, A3, A5 \\ \hline
		\begin{tabular}[c]{@{}l@{}} TWFE model   \\ (Section \ref{more groups and time regression model})     \end{tabular}  & None                                                                   & \begin{tabular}[c]{@{}l@{}}$\widehat{\beta}_1'$ from $E[Y_t| G, T] = \beta_0' + \beta_1'\textbf{I}(expanded_t)$\\ $ + \beta_2'\textbf{I}(G = 2014) + \beta_3'\textbf{I}(T = 2014)$\\ $ + \beta_4'\textbf{I}(G = 2015) + \beta_5'\textbf{I}(t = 2015)$\end{tabular}        & A1, A2, A6                                                                                                  \\ \hline
	\end{tabular}
\label{tab:Table 5}
\end{table}

\newpage 
\section{Bibliography}
\bibliography{refs}%

\newpage

\section{Appendix}\label{appendix}

\subsection{Identification of ATT for two groups in two time periods in general notation (from Section \ref{2x2 did})}\label{3.1}

\noindent \noindent With assumptions A1 and A2, we can identify $ATT(g,t)$ with observable data as follows:

\begin{align*}
ATT(g,t)  & = E[Y_{t}^{1;g} - Y_{t}^{0;g} | G = g]\\
& = E[Y_{t}^{1;g}| G=g] - E[Y_{t}^{0;g}  | G =g]\\
&= E[Y_{t}| G=g] - E[Y_{t}^{0;g} | G =g]\\
&= E[Y_{t} | G = g] - E[Y_{g-1}^{0;g} |G=g] - E[Y_{t}^{0;g} | G =g] + E[Y_{g-1}^{0;g} |G=g]\\
&= E[Y_{t} | G = g] - E[Y_{g-1}|G=g] - E[Y_{t}^{0;g}  | G =g] + E[Y_{g-1}^{0;g} | G=g]\\
&= E[Y_{t} - Y_{g-1}| G = g] - E[Y_{t}^{0;g}  - Y_{g-1}^{0;g} | G =g]\\
&= E[Y_{t} - Y_{g-1}| G = g] - E[Y_{t}^{0;g}  - Y_{g-1}^{0;g} | G > g]\\
&= E[Y_{t} - Y_{g-1}| G= g] - E[Y_{t} - Y_{g-1}| G > g] \\
\end{align*}

\subsection{Identification of ATT with two groups and two periods using linear regression modeling approach in general notation (from Section \ref{2x2 regression model})}\label{4.1}

\begin{align}\label{2x2 regression model formula app}
   E[\mbox{Y}_{t}|G, T] = \beta_0 +  \beta_1 \textbf{I}\{G = g\} + \beta_2 \textbf{I}\{T = t\}+ \beta_3\textbf{I}\{G = g\} \times \textbf{I}\{T = t\} 
\end{align} 

\noindent With model (\ref{2x2 regression model formula app}) and assumptions A1 and A2:

\begin{align*}
\beta_3 &= \left(\left(\beta_0 + \beta_1 + \beta_2 + \beta_3\right) - \left(\beta_0 + \beta_1 \right)\right) - \left(\left(\beta_0 + \beta_2\right) - \beta_0 \right)\\
&=  \left(E[Y_{t}| G = g] -E[Y_{g-1} | G = g]\right) - \left(E[Y_{t}| G > g] -E[ Y_{g-1} | G > g]\right)\\
&=  E[Y_{t} - Y_{g-1} | G = g] - E[Y_{t} - Y_{g-1} | G > g]\\
&= E[Y_{t}^{1;g} - Y_{t}^{0;g} | G = g]\\
&= ATT(g,t)
\end{align*}

\subsection{Identification of ATT with two groups and two time periods with covariates in general notation (from Section \ref{2x2 covariates did})}\label{3.2}

 \noindent One possible identification of $ATT(g,t)$ incorporating covariates and using assumption A3 and A4 is as follows

\begin{align*}
ATT(g,t)&=  E[Y_{t}^{1;g} - Y_{t}^{0;g} | G = g] \notag \\
&= E[Y_{t} - Y_{g-1}| G = g] - E[Y_{t}^{0;g}  - Y_{g-1}^{0;g} | G =g] \notag \\
&= E[Y_{t} - Y_{g-1}| G = g] - E\left[E[Y_{t}^{0;g}  - Y_{g-1}^{0;g}| \boldsymbol{X} =\boldsymbol{x}, G = g] | G = g\right] \notag   \\
&= E[Y_{t} - Y_{g-1}| G = g] - E\left[E[Y_{t}^{0;g}  - Y_{g-1}^{0;g}| \boldsymbol{X} =\boldsymbol{x}, G > g] | G = g\right] \notag  \\
&= E[Y_{t} - Y_{g-1}| G = g] - E\left[E[Y_{t}  - Y_{g-1}| \boldsymbol{X} =\boldsymbol{x}, G > g] | G = g\right] \notag  \\
&= E[Y_{t} - Y_{g-1}| G = g] - \left(E\left[E[Y_{t} | \boldsymbol{X} =\boldsymbol{x}, G > g\right]  - E\left[Y_{g-1}| \boldsymbol{X} =\boldsymbol{x}, G > g] | G = g]\right]\right)
\end{align*}

\subsection{Identification of ATT with two groups and two periods and covariates using linear regression modeling approach in general notation (from Section \ref{2x2 covariates regression model})}\label{4.2}

Consider a time-fixed variable $\boldsymbol{X}$ and linear model 

\begin{align} \label{2x2 covariate regression model formula app}
E[\mbox{Y}_{t}| G, T, \boldsymbol{X}] &= \beta_0^{*} +  \beta_1^{*} \textbf{I}\{G = g\} + \beta_2^{*} \textbf{I}\{T = t\}+ \beta_3^{*}\textbf{I}\{G = g\} \times \textbf{I}\{T = t\} + \beta_4^{*}\boldsymbol{X}.
\end{align}

\noindent With model (\ref{2x2 covariate regression model formula app}) and assumptions A1, A2, and A5:

\begin{align*}
ATT(g,t) &= E[Y_{t}^{1;g} - Y_{t}^{0;g} | G = g]\\
&=E\left[E[Y_{t}^{1;g} - Y_{t}^{0;g} | G = g, \boldsymbol{X}]| G = g\right]\\
&=E\left[E[Y_{t} - Y_{g-1} | G = g, \boldsymbol{X}] - E[Y_{t}^{0;g} - Y_{g-1}^{0;g} | G = g, \boldsymbol{X}]| G = g\right]\\
&=E\left[E[Y_{t} - Y_{g-1} | G = g] - E[Y_{t}^{0;g} - Y_{g-1}^{0;g} | G = g]| G = g\right]\\
&=E\left[E[Y_{t} - Y_{g-1} | G = g] - E[Y_{t}^{0;g} - Y_{g-1}^{0;g} | G > g]| G = g\right]\\
&=E\left[E[Y_{t} - Y_{g-1} | G = g] - E[Y_{t} - Y_{g-1} | G > g]| G = g\right]\\
&=E[\beta_3^* | G = g]\\
&= \beta_3^*
\end{align*}

\noindent Consider a time-fixed variable $\boldsymbol{X}$ interacted with time in linear model 

\begin{align} \label{2x2 tv covariate regression model formula app}
E[\mbox{Y}_{t}| G, T, \boldsymbol{X}] &= \beta_0^{**} +  \beta_1^{**} \textbf{I}\{G = g\} + \beta_2^{**} \textbf{I}\{T = t\}+ \beta_3^{**}\textbf{I}\{G = g\} \times \textbf{I}\{T = t\}  + \beta_4^{**}\boldsymbol{X} + \beta_5^{**}\boldsymbol{X} \times \textbf{I}\{T = t\}
\end{align}

\noindent With model (\ref{2x2 tv covariate regression model formula app}) and assumptions A1, A3, and A5:

\begin{align*}
ATT(g,t) &=E[Y_{t}^{1;g} - Y_{t}^{0;g} | G = g]\\
&=E\left[E[Y_{t}^{1;g} - Y_{t}^{0;g} | G = g, \boldsymbol{X}]| G = g\right]\\
&=E\left[E[Y_{t} - Y_{g-1} | G = g, \boldsymbol{X}] - E[Y_{t}^{0;g} - Y_{g-1}^{0;g} | G = g, \boldsymbol{X}]| G = g\right]\\
&=E\left[E[Y_{t} - Y_{g-1} | G = g, \boldsymbol{X}] - E[Y_{t}^{0;g} - Y_{g-1}^{0;g} | G > g, \boldsymbol{X}]| G = g\right]\\
&=E\left[E[Y_{t} - Y_{g-1} | G = g, \boldsymbol{X}] - E[Y_{t} - Y_{g-1} | G > g,\boldsymbol{X}]| G = g\right]\\
&= E\left[\left((\beta_0^{**} + \beta_1^{**} + \beta_2^{**} + \beta_3^{**} + \beta_4^{**}\boldsymbol{X} + \beta_5^{**}\boldsymbol{X}) - (\beta_0^{**} + \beta_1^{**}   + \beta_4^{**}\boldsymbol{X}) \right)\right.\\
&\hspace{0.5 cm}- \left.\left((\beta_0^{**} + \beta_2^{**} + \beta_4^{**}\boldsymbol{X} + \beta_5^{**}\boldsymbol{X}) - (\beta_0^{**} +  \beta_4^{**}\boldsymbol{X}) \right)| G = g\right]\\
&= E[\beta_3^{**}| G = g]\\
&= \beta_3^{**}
\end{align*} 

\subsection{Identification of ATT from regression model including age, sex, race, and region (all interacted with time) presented in last row in Table 1 in Section \ref{2x2 covariates regression model}}\label{app table1}

Model:

\begin{align} \label{2x2 big model app}
 E[Y_t|G, T, \boldsymbol{X}] &= \beta_0 + \beta_1\textbf{I}\{T = 2014\} + \beta_2\textbf{I}\{G = 2014\}  + \beta_3\textbf{I}\{T = 2014\} \times \textbf{I}\{G = 2014\} + \beta_4age\\
 &+ \beta_5age\times\textbf{I}\{T = 2014\}+ \beta_6\textbf{I}\{Male\} + \beta_7\textbf{I}\{Male\}\times\textbf{I}\{T = 2014\} + \beta_8\textbf{I}\{\text{Black}\} + \beta_9\textbf{I}\{\text{Black}\}\times\textbf{I}\{T = 2014\} \notag \\
 &+ \beta_{10}\textbf{I}\{SO\} \notag + \beta_{11}\textbf{I}\{SO\}\times\textbf{I}\{T = 2014\} +\beta_{12}\textbf{I}\{NE\} + \beta_{13}\textbf{I}\{NE\}\times\textbf{I}\{T = 2014\}+\beta_{14}\textbf{I}\{MW\} \notag\\
 &+ \beta_{15}\textbf{I}\{MW\}\times\textbf{I}\{T = 2014\} \notag
\end{align}\\

\noindent Components of $ATT(2014,2014)$:\\

\begin{align*}
E[Y_{2014}| G = 2014, \boldsymbol{X}] &= \beta_0 + \beta_1 + \beta_2 + \beta_3 + \beta_4 age + \beta_5 age + \beta_6 \mathbf{I}\{Male\} + \beta_7\mathbf{I}\{Male\} + \beta_8\mathbf{I}\{\text{Black}\} + \beta_9 \mathbf{I}\{\text{Black}\}\\
&+ \beta_{10} \mathbf{I}\{SO\} + \beta_{11} \mathbf{I}\{SO\}+ \beta_{12} \mathbf{I}\{NE\} + \beta_{13} \mathbf{I}\{NE\} + \beta_{14}\mathbf{I}\{MW\} + \beta_{15} \mathbf{I}\{MW\}
\end{align*}

\begin{align*}
E[Y_{2013}| G = 2014, \boldsymbol{X}] &= \beta_0 + \beta_1  + \beta_4 age +  \beta_6 \mathbf{I}\{Male\}+  \beta_8\mathbf{I}\{\text{Black}\} + \beta_{10}\mathbf{I}\{SO\} + \beta_{12}\mathbf{I}\{NE\} + \beta_{14}\mathbf{I}\{MW\}
\end{align*}

\begin{align*}
E[Y_{2014}| G > 2014, \boldsymbol{X}] &= \beta_0  + \beta_2  + \beta_4 age + \beta_5 age + \beta_6 \mathbf{I}\{Male\} + \beta_7 \mathbf{I}\{Male\} + \beta_8\mathbf{I}\{\text{Black}\} + \beta_9 \mathbf{I}\{\text{Black}\}\\
&+ \beta_{10} \mathbf{I}\{SO\} + \beta_{11}\mathbf{I}\{SO\}+ \beta_{12} \mathbf{I}\{NE\} + \beta_{13}\mathbf{I}\{NE\} + \beta_{14}\mathbf{I}\{MW\}+ \beta_{15}\mathbf{I}\{MW\}
\end{align*}

\begin{align*}
E[Y_{2013}| G = 2014, \boldsymbol{X}] &= \beta_0   + \beta_4 age +  \beta_6\mathbf{I}\{Male\}+  \beta_8\mathbf{I}\{\text{Black}\} + \beta_{10}\mathbf{I}\{SO\} + \beta_{12}\mathbf{I}\{NE\} + \beta_{14}\mathbf{I}\{MW\}
\end{align*}

\noindent With assumptions A1, A3, and A5:

\begin{align*}
ATT(2014,2014) &=E[Y_{2014}^{1;2014} - Y_{2014}^{0;2014} | G = 2014]\\
&=E\left[E[Y_{2014}^{1;2014} - Y_{2014}^{0;2014} | G = 2014, \mathbf{X}]| G = 2014\right]\\
&=E\left[E[Y_{2014} - Y_{2013} | G = 2014, \mathbf{X}] - E[Y_{2014}^{0;2014} - Y_{2013}^{0;2014} | G = 2014, \mathbf{X}]| G = 2014\right]\\
&=E\left[E[Y_{2014} - Y_{2013} | G = 2014, \mathbf{X}] - E[Y_{2014}^{0;2014} - Y_{2013}^{0;2014} | G > 2014,\mathbf{X}]| G = 2014\right]\\
&=E\left[E[Y_{2014} - Y_{2013} | G = 2014, \mathbf{X}] - E[Y_{2014} - Y_{2013} | G > 2014,\boldsymbol{X}]| G = 2014\right]\\
&= E\left[\left( \beta_2 + \beta_3 + \beta_5 age + \beta_7 \mathbf{I}\{Male\} + \beta_9 \mathbf{I}\{\text{Black}\} + \beta_{11} \mathbf{I}\{SO\} + \beta_{13} \mathbf{I}\{NE\} + \beta_{15} \mathbf{I}\{MW\} \right)\right.\\ 
&\hspace{0.5 cm} - \left. \left( \beta_2 + \beta_5 age + \beta_7 \mathbf{I}\{Male\} + \beta_9 \mathbf{I}\{\text{Black}\} + \beta_{11} \mathbf{I}\{SO\} + \beta_{13} \mathbf{I}\{NE\} + \beta_{15} \mathbf{I}\{MW\}\right)  | G = 2014\right]\\
&= E[\beta_3| G = 2014]\\
&= \beta_3
\end{align*} \\

\noindent We note that model (\ref{2x2 big model app}) does not adjust for age using splines with 3 knots, as the model in Table \ref{tab:Table 4.2} does. Identification of $ATT(2014,2014)$ with splines results in the same value of $\beta_3$, with each corresponding age term canceling out.

\subsection{Results of partial test of conditional parallel trends assumption (from Section \ref{more groups and time did})}\label{parallel trends test}

		\begin{figure}[ht]
			\centering
			\includegraphics[width = 11 cm]{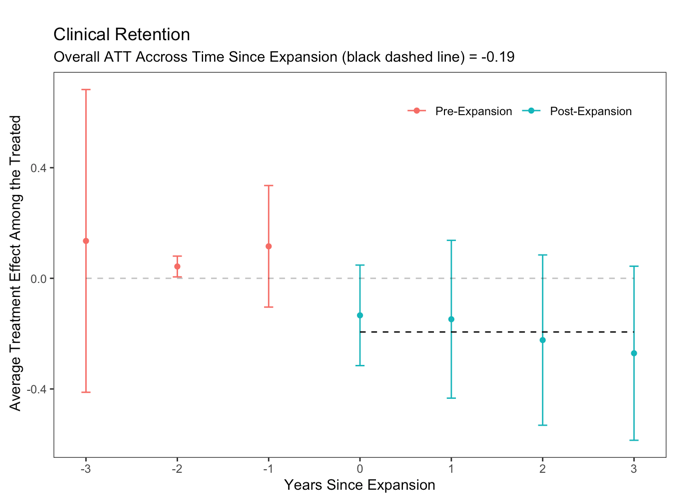}
			\caption{Pre-expansion and post-expansion values of $ATT(g, g+w)$ using method described in Section \ref{more groups and time did}. Pre-expansion estimates (red) are used to partially test parallel trends assumption. P-value for partial test of parallel trends: 0.91.}
   \label{fig:did parallel trends test}
		\end{figure}

\newpage

\end{document}